\documentclass[12pt,preprint]{aastex}

\slugcomment{Accepted for ApJ}
\shorttitle{Parallaxes for Two Subdwarfs}
\shortauthors{Dahn et al.}

\begin{document}

\title{Trigonometric Parallaxes for Two Late-Type Subdwarfs:\\
   LSR1425+71 (sdM8.0) and the Binary LSR1610$-$00 (sd?M6pec)}

\author{Conard C. Dahn, Hugh C. Harris, Stephen E. Levine, Trudy Tilleman,\\
        Alice K. B. Monet, Ronald C. Stone,\altaffilmark{1}
        ~Harry H. Guetter, Blaise Canzian,\altaffilmark{2}\\
        Jeffrey R. Pier}
\affil{US Naval Observatory, Flagstaff Station,
       10391 West Naval Observatory Road,\\
       Flagstaff, AZ 86001-8521}
\email{dahn@nofs.navy.mil}

\author{William I. Hartkopf}
\affil{US Naval Observatory, 3450 Massachusetts Ave., N.W.,
       Washington, DC 20392-5420}
\email{wih@usno.navy.mil}

\author{James Liebert}
\affil{Steward Observatory, University of Arizona,
       933 North Cherry Ave.,  Tucson AZ 85721}
\email{liebert@jayhawk.as.arizona.edu}

\author{Michael Cushing}
\affil{Institute for Astronomy, 2680 Woodlawn Drive,
       Honolulu, HI 96822}
\email{mcushing@ifa.hawaii.edu}

\altaffiltext{1}{Deceased 10 Sept. 2005} 
\altaffiltext{2}{Current address: L--3
        Communications/Brashear, 615 Epsilon Drive, Pittsburgh, PA 15238}

\begin{abstract}

Trigonometric parallax astrometry and BVI photometry are presented for two late-type
subdwarf candidates, LSR1425+71 (sdM8.0) and LSR1610$-$00 (sd?M6pec).  For the former
we measure an absolute parallax of 13.37~$\pm$~0.51 mas yielding M$_{\rm V}$~=~
15.25~$\pm$~0.09.  The astrometry for LSR1610$-$00 shows that this object is an
astrometric binary with a period of 1.66$\,\pm\,$0.01 yr.  The photocentric orbit is
derived from the data;  it has a moderate eccentricity (e$\,\approx\,$0.44$\,\pm\,$0.02)
and a semi-major axis of 0.28$\,\pm\,$0.01 AU based on our measured absolute parallax
of 31.02$\,\pm\,$0.26 mas.
Our radial velocity measure of $-$108.1$\,\pm\,$1.6 km s$^{-1}$ for LSR1610$-$00
at epoch 2006.179, when coupled with the observation of $-$95$\,\pm\,$1 km s$^{-1}$
at epoch 2005.167 by Reiners \&\ Basri, indicates a systemic radial velocity of
$-$101$\,\pm\,$1 km s$^{-1}$ for the LSR1610$-$00AB pair.  The galactic velocity components
for LSR1425+71 and LSR1610$-$00AB -- 
(U,$\,$V,$\,$W)$\,=\,$(84$\,\pm\,6$,$\;-$202$\,\pm\,$13,$\;$66$\,\pm\,$14) km s$^{-1}$  
and (U,$\,$V,$\,$W)$\,=\,$(36$\,\pm\,2$,$\;-$232$\,\pm\,$2,$\;-$61$\,\pm\,$2) km s$^{-1}$,
respectively.  For both stars, the velocities are characteristic of halo population
kinematics.  However, modelling shows that both stars have orbits around the galaxy
with high eccentricity that pass remarkably close to the galactic center.
LSR1425+71 has a luminosity and colors consistent with its metal-poor subdwarf
spectral classification, while LSR1610$-$00 has a luminosity and most colors
indicative of being only mildly metal-poor, plus a uniquely red $B-V$ color.
The companion to LSR1610$-$00 must be a low-mass, substellar brown dwarf.
We speculate on the paradoxical nature of LSR1610$-$00 and possible sources of its
peculiarities.

\end{abstract}


\section{Introduction}

     Over the last decade, spectroscopic surveys targeting faint stars with large
proper motions have identified many subdwarfs with M spectral types.  These surveys
generally make use of the spectral classification scheme developed by Gizis (1997)
which employs measures of the CaH and TiO bands in the 6300--7100~\AA\ spectral
region to form the indices CaH1, CaH2, CaH3, and TiO as defined in Reid et al.
(1995).  The locations in the three diagrams for the CaH indices versus
the TiO index serve to separate the stars into three spectral metallicity classes
designated MV for dwarfs with [M/H]$\,\approx\,$0.0, sdM for the cool counterparts
to the classical sdF--sdG subdwarfs with [M/H]$\,\approx\,-$1.2$\,\pm\,$0.3, and
esdM for extreme subdwarfs with [M/H]$\,\approx\,-$2.0$\,\pm\,$0.5 (Gizis \&\
Reid 1997).  The numerical spectral subclass is then determined from calibrated
relations with both the CaH2 and CaH3 indices.  At the time that this classification
scheme was established, there was only one star with subclass type later than sdM4.5
known -- namely LHS377, which was assigned the type sdM7 ``by definition" (Gizis \&\ Harvin
2006).

    It has since become routine to consolidate the essential features of the Gizis
classification scheme into a single diagram consisting of [CaH2~+~CaH3] versus TiO5
(cf., L\'epine et al. 2004), retaining the revised divisions between the three
metallicity subclasses proposed by Burgasser \&\ Kirkpatrick (2006).  Further revision
and extension of the Gizis system have been proposed by L\'epine et al. (2007).
A fourth subclass designated ``usdM" indicating ``ultrasubdwarfs" was added to the
three-subclass scheme to recognize the relatively few M stars which apparently have
metallicities as low as or perhaps below [M/H]$\,=\,-$2.5.  A realignment of the
subclass separators
was put forward based on a new index, $\zeta_{\rm TiO/CaH}$, which measures the
calibrated TiO5 spectral index as a function of the [CaH2~+~CaH3] index and is
expressed as a ratio with the solar metallicity value.  The separators are chosen
so that they run parallel to the locii in the [CaH2~+~CaH3,~TiO5] plane for the
observed components of wide common proper motion binaries with different spectral
subtypes.  A proposed list of subdwarf spectral classification standard stars were
presented covering types K7.0 through M8.5 for each of the subclasses subdwarf,
extreme subdwarf, and ultrasubdwarf.  As further suggested by Burgasser et al.
(2005), one really needs to simultaneously utilize spectral
classification criteria from a wider range of wavelengths and low-resolution data
covering the range 0.7--2.5 $\mu$m as a possible way to proceed in the future.

    However, while the number of earlier type M subdwarfs recognized in the
greater solar neighborhood has grown to several hundred due to these survey
efforts, subdwarfs with types sdM6 and later remain relatively rare.  Even as
recently as this past year, the number of subdwarfs with types sdM7 or later
numbered only around 15 (Burgasser et al. 2007; L\'epine et al. 2007).
The discovery of one of these few -- LSR1425+7102, hereafter referred to as
LSR1425+71 -- was announced several years ago by L\'epine et al. (2003b) and assigned
a classification of sdM8.0 on the Gizis (1997) system, making it the coolest sdM star
identified at that time.  The similarity in overall continuum slope with LSR2000+3057
(M6.0V) over the $6000-9000\,$\AA\ spectral region was noted.  Employing the Baraffe
et al. (1997; hereafter BCAH97) ``NextGen'' model atmospheres for a metal-poor
0.09M$_{\sun}$ star with [Fe/H]=$-$1.3 to estimate the absolute magnitude,
they derived a ``conservative distance estimate of 65$\,\pm\,$15 pc.''  This distance,
when combined with their measured proper motion of 0.635 arcsec yr$^{-1}$ in a position
angle of 254.7 deg and their measured radial velocity of $-$65$\,\pm\,$20 km s$^{-1}$,
yielded space velocity components of (U,V,W)\footnote{
Throughout this paper, U is measured radially outward toward the galactic
anticenter.}
=($-$65$\pm$22, $-$177$\pm$35, +64$\pm$27)
km s$^{-1}$, consistent with halo membership.  Note that LSR1425+71 is also adopted as
the classification standard star for spectral type sdM8.0 by L\'epine et al. (2007).

    Later in 2003, the same investigators announced the discovery of LSR1610$-$0040,
hereafter referred to as LSR1610$-$00, and suggested that it might be an early-type
sdL subdwarf (L\'epine et al. 2003a; hereafter LRS03).
At that time the only other sdL candidate was 2MASS$\,0532+8246$ discovered by Burgasser
et al. (2003) from spectroscopic follow-up on 2MASS photometric candidates.  L\'epine
et al. (2003a) noted that while their optical spectrum of LSR1610$-$00
(coverage$\approx\,6000-10,000\,$\AA; resolution$\,\approx\,$7$\,$\AA) showed obvious
similarities to their spectrum of LSR1425+71, LSR1610$-$00 possessed a distinctly steeper
pseudocontinuum (implying a cooler temperature) and, more significantly, showed strong
Rb~I~$\,7800,7947\,$\AA$\;$ lines which are more typically present in L-type dwarfs
rather than M-dwarfs.  With its weak TiO bands and totally absent VO bands LSR1610$-$00 did
not fit anywhere in the known sequences of dwarf M or subdwarf M stars.  Again using
the BGAH97 models, they ``conservatively" estimated the distance to LSR1610$-$00 to be
16$\,\pm\,$4 pc.  Centroids of the Rb~I lines together with those from the K~I doublet
(7665,~7699$\,$\AA) yielded a heliocentric radial velocity of $-$130$\,\pm\,$20 km s$^{-1}$.
Using their measured proper motion of 1.46 arcsec yr$^{-1}$ in a position angle 212.0 deg
yields galactic space velocity components of (U,V,W)=($-$117$\pm$18, $-$108$\pm$24,
$-$10$\pm$19) km s$^{-1}$, suggesting that LSR1610$-$00 is most likely a member of the
old disk population.\footnote{Note that the U,~V,~W values presented in Table 1 of LRS03
are incorrect, even for the input values these authors employ.}

    Combining new moderate resolution spectroscopy covering 0.8--4.1$\,\mu$m with the
0.6--1.0$\,\mu$m data from LRS03, Cushing \&\ Vacca (2006; hereafter
CV06) made a comprehensive study of the peculiarities and contradictions presented in
the spectrum of LSR1610$-$00.  Similar spectral data for three other sdM stars (LHS~3409, sdM4.5;
LHS~1135, sdM6.5; LSR2036+5059, sdM7.5) and the M6.0V star GL406 were employed for comparison.
GL406 is the well-studied high proper motion star Wolf~359 which is a primary MV spectral
classification standard (Kirkpatrick et al. 1991) and a well established active
flare star (CN Leo).  Based on its (U,V,W) galactic kinematic components, GL406 is formally
a member of the old disk population (Leggett 1992).  However, Pavlenko et al. (2006) argue that
its age is most likely younger than 0.4 Gyr -- that is, younger than the Hyades cluster
(age$\,\sim\,$0.6 Gyr) which is often taken as the upper age limit for the young disk population.
CV06 concur with LRS03 that, in terms of spectral features, LSR1425+71 is a better match with LSR1610$-$00
than is GL406.  On the other hand, in terms of overall relative spectral
energy distribution (SED) over the entire wavelength range 0.6--4.1~$\mu$, LSR1610$-$00 and GL406 are
remarkably similar, despite the fact that some features in the 7640--8300~\AA\ spectral region
are noteably dissimilar, especially the stronger TiO bandhead and the much weaker
Rb~I doublet in Gliese 406.  Given all of the contradictory evidence summarized by CV06
(see Table 3), these authors conclude that LSR1610$-$00 is most likely a mildly metal-poor, mid-M
dwarf but further call the star ``schizophrenic" to acknowledge its several spectral peculiarities.
In particular we note the abnormally strong Al~I doublet at 1.313~$\mu$m which suggests that
aluminum is most likely overabundant compared with solar.

    In a study concurrent with that of CV06, Reiners \&\ Basri (2006; hereafter RB06) obtained
high-resolution (R$\,\approx\,$31000) spectra in several windows of the 0.7--1.0$\,\mu$m
region for LSR1610$-$00, GL406, and 2MASS~1439+1929 (hereafter, 2M1439+19).  The latter is a
L1V star with an established trigonometrically determined luminosity (Dahn et al. 2002).
Comparision of the strengths of TiO bands with heads at 7050, 8430, and 8870 \AA\ do not
preclude the possibility that LSR1610$-$00 might be slightly metal deficient.  On the other hand,
comparison of metal hydride bands (e.g., CaH at 6800~\AA\ and FeH at 9900~\AA) indicate
that LHS1610$-$00 can not be very much more metal-deficient than is GL406.  As pointed out earlier
by both LRS03 and CV06, the primary discrepancies arise for a few atomic lines.  In particular,
the Rb~I lines at 7947 and 7800 \AA\ agree much better with those in 2M1439+19, not only
in strength but in shape.  In contrast, the Cs~I lines at 8520 and 8944 \AA\ are much weaker
in LSR1610$-$00 than in 2M1439+19, with the redder one being almost undetectable.  Not only are the
CS~I lines more similar to those in GL406, the one at 8520 \AA\ is significantly stronger
than in GL406.  The strengths of the Ca~II triplet components at 8498 and 8542 \AA\ imply
that LSR1610$-$00 must be at least as warm as GL406.  Comparing the spectra of LSR1610$-$00 and 2M1439+19
with PHOENIX atmospheric model predictions (cf., Allard et al. 2001), RB06 found that the
observed Rb~I and Cs~I lines {\it could} be simultaneously reproduced at a satisfactory level
by a slightly metal-deficient model ([Fe/H]~$\approx$~$-$1) with T$_{\rm eff}$~=~2800~K, 
which alters the pseudo-continuum level appropriately for LSR1610$-$00.\footnote{L\'epine et al.
(2004) announced the discovery of LSR0822+1700 as a second subdwarf M star that showed both RbI
and CsI lines.  In this instance, the location in the [CaH2~+~CaH3,~TiO5] plane clearly indicates
low metallicity.  LSR0822+1700 has been designated as a usdM7.5 spectral classification standard
star by L\'epine et al. (2007).  USNO initiated a parallax determination for this star in March
2008.} Based on this result RB06
suggested a spectral type of sd?M6 would be appropriate.  However, given the remaining spectal
peculiarities not addressed by RB06, we prefer to adopt the spectral type designation of
sd?M6pec for the present.  Finally, RB06 measured a high-precision heliocentric radial
velocity for LSR1610$-$00 of $-$95$\,\pm\,$1 km s$^{-1}$ by cross-correlating its spectum with that
of GL406 in the spectral region around 8000$\,$\AA.  They discuss the discrepancy between
their result and that of LRS03 ($-$130$\,\pm\,$15 km s$^{-1}$) measured from much lower
resolution spectroscopy and suggest that the LRS03 value might require a corrective offset
of +20 km s$^{-1}$.

Calibration of the extension to the Gizis subdwarf-M spectral classification system to
physical parameters such as total luminosities and effective temperatures, such as proposed
by L\'epine et al. (2007), will require reliable trigonometrically determined
distances.  A representative sample of such M-type subdwarfs is currently being observed
at the Flagstaff Station.  Both LSR1425+71 and LSR1610$-$00 were added to the USNO CCD trigonometric
parallax program back in the spring of 2003 and reliable results are now available for
these two stars.  Since LSR1425+71 and LSR1610$-$00 have been discussed and compared with one
another repeatedly in the literature, we present our results for both stars together here.

\section{Observations}

The new observations presented here include: 1) parallax astrometry for both LSR1425+71
and LSR1610$-$00; 2) BVI photometry for both stars; and 3) a new radial velocity measure
for LSR1610$-$00.  The parallax astrometry was carried out with the 1.55~m Strand 
Astrometric Reflector at the Flagstaff Station of the US Naval Observatory using the
TEK2K CCD Camera.  The observational procedures followed were those summarized
previously in Dahn et al. (2002).  The astrometrically flat wide--I filter was employed
for both fields.  Photometry on the Johnson--Cousins system was obtained in the BVI
bandpasses for both fields using the USNO 1~m telscope and following standard procedures
(i.e., standard stars from Landolt 1992 observed each night; small color terms
determined and applied nightly).  Differential color refraction corrections for astrometric
observations taken slightly off the meridian were derived from the V$-$I colors of the
parallax star and the reference stars employed, as described in Monet et al. (1992).
The corrections from relative to absolute parallax were derived from photometric
parallaxes of the individual references stars using M$_{\rm V}$ versus V$-$I
relations calibrated with stars with large trigonometric parallaxes.  The new radial
velocity measure of LSR1610$-$00 was obtained using the University of
Arizona/Harvard--Smithsonian 6.5~m MMT reflector.
Details regarding the results follow.

\subsection{Astrometric and Photometric Results for LSR1425+71}

A total of 73 acceptable observations of this field have been obtained
since it was added to the active program in April 2003, providing a total epoch
coverage of over 4 years.  The absence of
suitable reference stars close to the parallax star has necessitated the use of a
fairly spread out frame which extends to about 3 arc minutes from the target star.
However, since the exposure times for this field are typically 15-20 minutes (depending
on seeing and transparency) good averaging over atmospheric fluctuations was achieved
resulting in formal rms single observational errors for the reference stars of
$\pm$~2.14 mas and $\pm$~2.25 mas in RA and DEC, respectively.

The astrometric and photometric results are presented in Table 1.  As we shall show in
Section 3 below, our parallax and photometry confirm the sudwarf status of LSR1425+71 in
three color--absolute magnitude diagrams.  Likewise, adopting the radial velocity of
$-$65$\,\pm\,$20 km s$^{-1}$ measured by Lepine et al. (2003b), the derived
U,~V,~W galactic kinematic components are entirely consistent with a halo population star,
but see Sec. 5 below.

\subsection{Astrometric and Photometric Results for LSR1610$-$00}

Astrometric observations of LSR1610$-$00 commenced in June 2003 and by early 2005
the residuals from the solution for parallax and proper motion clearly indicated 
that this object is an unresolved, astrometric binary.  Observational coverage
of the field was intensified at that time.  To date, a total of 219 acceptable
observations spanning an epoch range of over 4 years has been obtained on this
field.  Our results indicate that over two full periods of the astrometric orbit
have now been covered, permitting solutions for the photocentric orbital elements
as well as the parallax and proper motion for this binary system.

In carrying out the astrometric reductions for LSR1610$-$00 we have chosen to employ
an iterative procedure wherein the standard algorithms were first solved for parallax
and proper motion.  The residuals from this solution were then solved for the motion
of the center of light about the center of mass.  This resulted in preliminary orbital
elements for the photocenter.  These elements were then used to ``correct'' the originally
measured parallax star positions, after which the solution was rerun for parallax and
proper motion.  The residuals from this revised solution were then resubmitted to the
binary star reduction algorithms to obtain refined orbital elements.  This iterative
process was repeated until convergence had been achieved.  Only four iterations were
required.

The results for LSR1610$-$00AB are summarized in Table 2 and the derived orbital
parameters for the photocentric orbit are presented in Table 3.  Figure 1 shows the
the residuals in RA and DEC after both the parallax and proper motion of the
system have been removed, and the Table 3 orbit fit to these data.  In general, the
derived orbital elements appear to be quite robust.  The possible exception is the
eccentricity, because observations at phases around periastron passage are still not
as numerous as desirable.
Upcoming observations during the March--October 2008 interval should resolve this
situation since we predict that the photocenter will pass through eastern elongation
during this window.

We also note that since the observer's line of sight to the LSR1610$-$00AB system is only 
6.8$\,\pm\,$1.0 degrees out of the plane of the photocentric orbit, the question of
possible eclipses is raised.  However, as we shall conclude in Section 4, the radii
of the most probable components are small.  No evidence of eclipses can be seen
in our astrometric observations, but our time coverage during anticipated close
passages is too spotty to provide any meaningful constraint.

Our confirmation that LSR1610$-$00 is indeed an unresolved binary system opens a new
dimension for understanding the seemingly ``schizophrenic spectrum'' of LSR1610$-$00
described by CV06.  However, as we shall discuss in Section 4, the published radial
velocities available for this star in the literature --
$-$130$\,\pm\,$20 km s$^{-1}$ at Epoch = 2003.137 by LRS03 and
$-$95$\,\pm\,$1 km s$^{-1}$ at Epoch = 2005.167 by RB06 -- present difficulties for
interpreting our photocentric orbit.  Hence, priority was given to obtaining a new,
independent measure of the radial velocity at an orbital phase different from those
of the earlier results.

\subsection{A New Radial Velocity Measure for LSR1610$-$00AB}

The radial velocity observation of LSR1610$-$00 was obtained on 2006 March 6 (U.T.)
using the 6.5~m MMT and red channel spectrograph.  Based on the astrometry for the
apparent orbit available at that time we anticipated that the photocenter of the
system would be near elongation during the spring of 2006.  The 1200 g/mm grating
blazed at 7700~\AA\ was employed yielding a resolution of 2.1~\AA\ with a one arc
second slit width.  The wavelength coverage used for the measurements was from about
7640~\AA\ (omitting the edge of the spectrum in the atmospheric ``A" band) to 8300~\AA.
The principal strong lines encompassed in this spectrum include the K~I resonance doublet
at 7665, 7699~\AA, the Rb~I doublet at 7800, 7947~\AA, and the subordinate Na~I 8183,
8194~\AA\ doublet.  GL406 was observed as a radial velocity standard for cross-correlation
reductions.  

The observations were generally obtained in pairs of 600 second and 300 second
integrations for LSR1610$-$00 and GL406, respectively.  Each pair was bracketed by
observations of the helium-neon-argon arc lamp at the position of the object.
Excellent wavelength fits were obtained after extractions of the two-dimensional
images to obtain object apertures, using standard IRAF tasks for this purpose.
Separate solutions for all arc~/~object pairings were always accurate to better
than $\pm\,$0.1~\AA, generally accurate to $\approx\,\pm\,$0.05~\AA.  Cross-correlations
were carried out between each of the six LSR1610$-$00 observations and each of the
eight observations of Wolf~359 using the IRAF task ``fxcor".

Difficulties with employing the cross-correlation approach were encountered due to
the extreme fringing of the chip at these wavelengths, resulting in 50--100~\AA\
``bumps" which could not be removed by the flatfields.  Attempts at filtering the
data in favor of the smaller spacings of the atomic lines did not alleviate these
difficulties.  Consequently, the cross-correlation technique was abandonded
and the radial velocity of LSR1610$-$00 was determined from line profile fittings to
four individual lines -- Rb~I (7800, 7947~\AA) and Na~I (8183, 8194~\AA) -- for each
of the six LSR1610$-$00 spectra.  The actual wavelength values (in air) employed for
these four lines (7800.268, 7947.603, 8183.256, and 8194.824~\AA, respectively) were
obtained from the current National Institute of Standards and Technology Atomic
Spectra Database Web site (NIST 2008).  Rejecting one of the six LSR1610$-$00 observations
as an obvious outlier, the remaining five observations yielded a measured radial
velocity of $-$135.7 km s$^{-1}$.  Applying a calculated correction of +27.6 km
s$^{-1}$ to convert this to heliocentric,  we obtain V$_{\rm rad}$ = 
$-$108.1$\,\pm\,$1.6 km s$^{-1}$ at Epoch = 2006.179.

Even allowing for reasonable orbital variation, the
LRS03 result is not compatable with the more recent measures determined from
significantly higher resolution spectra.  RB06 discussed the possible need for
a roughly +20 km s$^{-1}$ adjustment to the LRS03 value based on similar systematic
effects noted by Burgasser et al. (2003) in the case of radial velocity measures
for 2MASS~0532+8246.  However, the large formal uncertainty attached to the
LRS03 value alone renders this measure of little value in assessing the nature
of the LSR1610$-$00AB binary system.  Hence, we will disregard the LRS03 radial velocity
measure from further consideration.  Taking into account the orbital phases of the
RB06 determination and our new observation (cf. Figure 1), we conclude that the
systemic velocity of the LSR1610$-$00AB is approximately $-$101$\,\pm\,$1 km s$^{-1}$
and that the velocity semi-amplitude for the photocenter of the binary is
$\lesssim\,$7 km s$^{-1}$.  With this systemic velocity and the parallax and
proper motion, we calculate the U,~V,~W galactic kinematic components given in
Table 2.  They are indicative of a halo population star, but see Sec. 5 below.

\section{Locations in Color--Magnitude Diagrams}

With reliable trigonometrically determined distances now available for
both LSR1425+71 and LSR1610$-$00AB, these stars can be placed in a variety of
absolute magnitude versus color diagrams (CMDs) to assess these stars'
relationship to each other, and to field M-dwarfs and sdM stars in general.
Evolutionary stellar models should provide additional guidance toward
understanding the status of both stars and, in particular, for inferring
what viable components might comprise the LSR1610$-$00AB system.  Models
for stars with masses corresponding to dwarf M and L spectral types have
been developed by several groups over the past decade, most noteably those
by the Lyon collaboration (BCAH97; Baraffe
et al. 1998, hereafter BCAH98; Chabrier et al. 2000), by the University of
Arizona group (Burrows et al. 2001; Burrows et al. 2006),
and most recently by the collaborators modeling the HST ACS (Advanced Camera
for Surveys) photometric survey of Galactic globular clusters (Sarajedini
et al. 2007; Dotter et al. 2007).  While differing in a variety of details,
the gross properties of the results from these three efforts agree where
they overlap in terms of mass and metallicity.  Hence, we will restrict
comparisons below primarily with the Lyon models whose results are directly
accessible in terms of absolute magnitudes and colors.

In presenting the Lyon models, BCAH97 and BCAH98 gave particular attention
to three such diagrams -- M$_{\rm V}$ versus V$-$I, M$_{\rm K}$ versus
I$-$K, and M$_{\rm K}$ versus J$-$K.  These three CMDs are discussed in turn
in the following subsections.

\subsection{The M$_{\rm V}$ versus V$-$I Diagram}

The locations of LSR1425+71 and LSR1610$-$00AB in the observed M$_{\rm V}$ versus
V$-$I diagram are presented in Figure 2.  Included for reference are a 
selection of field dwarfs and subdwarfs which possess both (a) published
USNO trigonometric parallaxes, and (b) VI photometry determined at the
Flagstaff Station.  The parallax determinations can be found in
Harrington et al. (1993; and earlier papers cited therein), Monet et al.
(1992), Dahn et al. (2002), and Reid et al. (2003).

Where parallax determinations are also available from other observatories
(mainly for brighter stars), the weighted mean parallaxes as derived by
van Altena et al. (1995) were adopted.  The photometry
includes considerable unpublished VI measures carried out at the
Flagstaff Station over many years.  M-type dwarfs are plotted as
filled circles and 5 early L-type dwarfs are included as filled
triangles.  Dwarf stars are included in Figure 2 only if the formal
errors for the parallax and the V magnitude combine for a formal
uncertainty in M$_{\rm V}$ of $\pm\,$0.15 mag or less.  This restriction
results in enough data points both to map out the primary main sequence
and also to illustrate the real physical spread of stars in this region
of the diagram, presumably due to small variations in metallicity (Bonfils et
al. 2005) and
due to unresolved binary pairs.  Subdwarfs are represented by open
circles in Figure 2 and those shown are restricted to stars with formal
uncertainties in M$_{\rm V}$ of $\pm\,$0.45 mag or less.  All stars
are plotted with their formal errors in both coordinates displayed.
Three late-type dwarfs either known to be binaries (2M0149+29 and
2M0746+20) or suspected of being binary (2M0345+25) are flagged
with `d' and `d?', respectively.  Also labeled in Fig.2 are GL406 (M6.0V)
and LHS377 (sdM7), the lowest luminosity M-type subdwarf with a
trigonometrically determined distance (Monet et al. 1992) known to date.

Three conclusions can be drawn from the star locations in Figure 2 alone.
First, in terms of overall match in SEDs,
the similarities between LSR1610$-$00AB and GL406 noted by CV06 for wavelengths
longer than 0.6~$\mu$m must extend at least down through wavelengths
corresponding to the V bandpass -- that is, at least to 0.5~$\mu$m.
Furthermore, the similarities here
established are for {\it absolute} energy distributions, not merely
{\it relative} ones.  Second, the components comprising LSR1610$-$00AB can not
include objects like either LSR1425+71 or LHS377 since both of these stars
are significantly more luminous in M$_{\rm V}$ and bluer than the combined
light from the astrometric binary system.  Finally, it is unlikely that
one can invoke a normal, early-type dL star as the B component of the
LSR1610$-$00AB system in order to explain the strong Rb~I$\,7800,7947\,$\AA$\;$
lines seen in the L\'epine et al. (2003a) optical spectrum.  In the
I-bandpass, even the earliest L dwarfs are approximately 2 magnitudes fainter
than what would be required for an A component satisfying the location of
the combined light in Fig. 2.  Such a large magnitude difference would greatly
reduce the visibility of the Rb~I lines, most likely rendering them undetectable.

Models with
solar (or near solar) metallicity continue to present problems, especially
for wavelengths shortward of $\sim$~1.0~$\mu$m where important molecular
opacity sources are still not included or are incomplete.  Some aspects of
the current status of such models are illustrate in Figure 2 by the solid
black line which represents the NextGen (BCAH98) isochrone for stars with
masses $\ge$~0.1~M$_{\sun}$ for an evolved age of 0.5 Gyr and by the solid
red line which represents the DUSTY (Chabrier et al. 2000) isochrone for
stars with masses $\le$~0.1~M$_{\sun}$, also for an evolved age of 0.5 Gyr.
For M$_{\rm V}\;<\;$19 (T$_{\rm eff}\;>2800\;$K) dust formation is unimportant
implying that the offset between the NextGen and DUSTY models is primarily
due to improved treatment of atmospheric opacity for TiO and, to a lesser
extent, CaH (Allard et al. 2000).  Still, the models remain
too blue by 0.1--0.2 mag in the region of Figure 2 corresponding to LSR1610$-$00AB
and GL406.  While better agreement between the DUSTY models and the field dM
stars is realized for a model evolutionary age of 0.1 Gyr (see Chabrier et.al
200, Fig.~4, and the dotted red line in the present Fig.~2.), such a young age
(younger than the Pleiades cluster!) is certainly unrealistic for the mixture
of young/old disk population field M-dwarfs in the solar neighborhood (Leggett 1992).

The blue lines in Figure 2 represent the BCAH97 isochrones for metal deficient
stars with [M/H] = $-$1.0, $-$1.5, and $-$2.0 and ages $\ge$~5 Gyr.  These
models appeared to be in satisfactory agreement with both the HST observations
of three globular clusters (M15, NGC6397, $\omega$~Cen) and field halo subdwarfs
with trigonometrically determined distance.  All of the aforementioned data seemed
to be well represented by isochrones with metallicities [M/H]~$\sim$~$-$1.3
to $-$1.5, corresponding to [Fe/H]~$-$1.6 to $-$1.8 when allowance for
oxygen enhancement over the evolutionary age of the Galaxy is taken into account.
This agreement provided hope that the BCAH97 models
might be reliable for detailed comparison with observations of metal-deficient
stars where the importance of double-metal molecular bands (e.g., TiO, VO) are
expected to be less important.  However, at the time of the BCAH97 paper, the
globular cluster observations extended only down to luminosities of M$_{\rm V}$~$\sim$~12
and the observed field subdwarfs only reached down to M$_{\rm V}$~$\sim$~14.5.
Figure 2 shows that model predictions for $-$2.0~$\le$~[M/H]~$\le-$1.0 effectively
converge for the range of interest for LSR1425+71 and LSR1610$-$00AB, that is for
M$_{\rm V}$ between $\sim$~15.0 and $\sim$~17.0.  We note that both of the subdwarfs
LSR1425+71 and LHS377 lie roughly 0.2 mag above the model locii (or alternately,
about 0.1 mag redward of the models).

The less than satisfactory agreement between the observations of subdwarfs with
types of early-M or later and current metal-deficient models in the M$_{\rm V}$
versus V$-$I diagram is almost certainly a continuing problem with the models.
The recent study of NGC6397 by Richer et al. (2007) demonstrated excellent agreement
between models appropriate for the observed distance ($\approx$~2.6~kpc), metallicity
([Fe/H]~$\approx$~$-$2.0; [$\alpha$/Fe]~$\approx$+0.3), and age ($\approx$~12~Gyr)
of this cluster and the HST/ACS observed CMD for over 11 magnitudes of luminosity
extending well into the giant branch, down through the cluster turnoff region, and
for over 8 magnitudes of the cluster main-sequence.  However, below a luminosity
corresponding to M$_{\rm V}$~$\sim$~13 the same systematic discrepancy between the
models and observations is seen for NGC6397 that we find in Figure 2.  This 
undoubtedly reflects incompleteness in the model atmospheric opacities,
especially at wavelengths shortward of $\sim$~1.0~$\mu$m.  Consequently, we still
must remain cautious when drawing detailed conclusions based on the locations of
M-type dwarfs and subdwarfs in the M$_{\rm V}$ versus V$-$I CMD.

\subsection{The M$_{\rm K_{\rm s}}$ versus I$-$K$_{\rm s}$ Diagram}

Less influence due to incomplete molecular opacity sources in the atmospheric model
calculations is anticipated for infrared bandpasses.  BCAH98 discussed the M$_{\rm K}$
versus I$-$K CMD, pointing out that it provided a ``powerful diagnostic for metallicity.''
Figure 3 shows locations of LSR1425+71, LSR1610$-$00AB, LHS377, and GL406 in the 
M$_{\rm K_{\rm s}}$ versus I$-$K$_{\rm s}$.  Here we are using photometry extracted
from the 2MASS point source catalog and retain the subscript ``s'' to
distinguish the 2MASS short-K bandpass from the CIT bandpass.  The filled and open circles
are field dwarfs and subdwarfs, respectively, as in Figure 2.  (Since the locii of dwarfs
and subdwarfs merge in this CMD, and form a nearly vertical sequence, we only plot field 
stars for M$_{\rm K_{\rm s}}$~$\ge$~7.3.)  Also included in Figure 3 is 2MASS~0532+8246
(hereafter, 2M0532+82), the first L-type subdwarf discovered by Burgasser et al. (2003).
A trigonometric parallax determined from ASTROCAM near-infrared observations carried out
at the USNO Flagstaff Station places the star at a distance of 26.7~$\pm$~1.2 pc and
yields clear halo kinematic properties (Burgasser et al. 2008).  Spectroscopically,
2M0532+82 is classified as sdL7 (Burgasser et al. 2007).  (The adopted
I magnitude used to plot 2M0532+82 in Figure 3 comes from Burgasser et al. 2008.  Note
that this star could not be included in Figure 2 since a V magnitude is still unavailable.)
The model locii included in Figure 3 are the same as those shown in Figure 2, except that
we include here the BCAH97 [M/H]~=~$-$1.3 isochrone for ages $\ge$~5 Gyr, in addition to
those for [M/H]~=~$-$1.0,~$-$1.5, and $-$2.0, since they are well separated in this CMD.
The transformation relations derived by Carpenter (2001) have been applied to the BCAH97
tabulated values to convert them from the CTI photometric system to the 2MASS system
employed here.

Some results to be noted from Figure 3 include the following.  First, the field subdwarfs
seem to be reasonably in accord with the BCAH97 metal-deficient isochrones for
M$_{\rm K_{\rm s}}$ below about 8.5.  However, for M$_{\rm K_{\rm s}}$~$<$~8.0 the 
isochrones run well to the blue side of the few observational points available.  LSR1425+71
is located nearly on the model locus for [M/H]~=~$-$1.3, consistent with its being a
later-type counterpart to the field sdF/sdG stars.  It is interesting to note that LSR1425+71
is only about 0.3 mag less luminous in M$_{\rm K_{\rm s}}$ than the two field subdwarfs
LHS~1742a (M$_{\rm K_{\rm s}}$~=~9.74, I$-$K$_{\rm s}$~=~1.96) and LHS~3061
(M$_{\rm K_{\rm s}}$~=~9.65, I$-$K$_{\rm s}$~=~1.99).  This contrasts to the relative
locations of these three stars in Figure 2, where LHS~1742a is about 0.8 mag more luminous
than LSR1425+71 in M$_{\rm V}$ and LHS~3061 is a full magnitude more luminous than LSR1425+71.
Second, The relative locations of LSR1425+71 and LHS377 in Figure 3 indicate that LHS377 is
significantly less metal deficient than LSR1425+71.  Accepting the models at face value might
lead to an estimate of [M/H]~$\approx$~$-$0.8 for LHS377.  Third, the location of LSR1610$-$00AB
in the midst of the disk field dwarfs again offers no clue to the nature of the individual
components making up this binary.  That is, it does not appear to be overluminous as would
be the case if both components of the binary were contributing significant light to the
photocenter.  And finally, the sdL7 2M0532+82 lies alone in the lower part of Figure 4.
Since the BCAH97 models include masses down to only 0.083~M$_{\sun}$, no reliable inference
concerning the metallicity of 2M0532+82 is possible from Figure 4.  Burgasser et al. (2008)
made a ``best guess" estimate of [M/H]~=~$-$1, implying a mass of 0.0783~$\pm$~0.0013 M$_{\sun}$
for this unique object.

As was the case in Figure 2, the DUSTY model isochrone for an age of 0.5 Gyr (solid red line)
does a poor job in representing the field dwarfs.  Likewise, the DUSTY model isochrone for
an age of 0.1 Gyr (dotted red line), while in better agreement with the observational points,
is still not fully satisfactory.

\subsection{The M$_{\rm K_{\rm s}}$ versus J$-$K$_{\rm s}$ Diagram}

If the discrepancies noted in the previous two CMDs between observations of field M-dwarfs
and solar metallicity model isochrones is primarily due to incompletenesses in the
atmospheric opacities for wavelengths blueward of $\sim$~1.0~$\mu$m, improved agreement
might be expected for the M$_{\rm K_{\rm s}}$ versus J$-$K$_{\rm s}$ CMD.  Figure 4
shows the stars under consideration in that CMD where the field dwarfs and subdwarfs, as
well as model isochrones included, are the same as in Figure 3.  As before, the model data
points have been converted from the CTI photometric system to the 2MASS system via the
Carpenter (2001) transformations.  While shape-wise the locii for both DUSTY model
isochrones (red solid and dotted lines) better reflect the trend of the M-dwarf observations
to branch off to redder J$-$K$_{\rm s}$ below M$_{\rm K_{\rm s}}$~$\approx$~10.0, the 0.5 Gyr
isochrone (solid red line) fails to reproduce the field M-dwarf observations at all colors
and luminosites plotted here.  Even the 0.1 Gyr isochrone (dotted red line) fails to
provide a satisfactory fit.

The field subdwarfs, including both LSR1425+71 and LHS377, seem to be well reproduced by the
BCAH97 models in Figure 4.  Again, accepting the fits at face value would imply metallicities
of [M/H]~$\approx$~$-$1.5~$\pm$~0.3 and [M/H]~$\approx$~$-$1.0~$\pm$~0.2 for LSR1425+71 and
LHS377, respectively.  The location of LSR1610$-$00AB amongst the field M-dwarfs in a region of
the diagram where the trend of the observational data points is nearly vertical prevents us
from inferring much of interest about possible components for the system.  We can note,
however, that there is nothing to suggest that LSR1610$-$00AB is a subdwarf.

In summary then, the CMDs discussed here all seem to support LSR1425+71 being a later-type
counterpart to the classical sdF/sdG field subdwarfs while LSR1610$-$00AB exhibits no subdwarf
characteristics.  Both the field M-dwarfs and the field M-subdwarfs remain inadequately
reproduced by current models in the M$_{\rm V}$ versus V$-$I CMD.  The field M-subdwarfs
seem to be reproduced satisfactorily in both the M$_{\rm K_{\rm s}}$ versus I$-$K$_{\rm s}$ 
and the M$_{\rm K_{\rm s}}$ versus J$-$K$_{\rm s}$ CMDs.  The solar metallicity and
near-solar metallicity models fail to adequately represent the field star M-dwarf data
points in all threee CMDs.  Hence, no substantive inferences can be drawn about the
components comprising the LSR1610$-$00AB binary from these diagrams.

\subsection{The B$-$V versus V$-$I Diagram}

The similarities in SEDs between GL406 and LSR1610$-$00AB noted in the above three CMDs do not,
apparently, extend to wavelengths as short as the B bandpass.  Figure 5 shows the location of
the stars in the B$-$V versus V$-$I color--color diagram.  Unfortunately, there
are only a limited number of late-type parallax dwarfs and subdwarfs with
measured B magnitudes, since they are generally very faint at shorter wavelengths. 
The filled and open circles in Figure 5 are the dwarfs and subdwarfs, respectively,
shown in Figure 2 where B$-$V photometry is available.  The open triangles are
stars spectoscopically classified as subdwarfs with Flagstaff Station
photometry but lacking suitably precise parallax distance determinations at
this time.  As is well recognized (e.g., Dahn et al. 1995; Fig. 3), subdwarfs
usually lie above the dwarfs in B$-$V versus V$-$I and by amounts which appear
to correlate with metal deficiency.  Figure 5 indicates that
the blue flux from the atmospheres of LSR1610$-$00AB is highly suppressed when
compared with GL406.  Lacking spectroscopic coverage over the blue region at
this time, we can only speculate about possible causes (see Sec. 6 below).  

\section{Inferring Plausible Components for the LSR1610$-$00AB System}

The observed orbit given in Table 3 is that of the photocenter (the combined
light) of the two components.  The photocenter motion is identical to the
motion of the primary A component if the secondary star contributes negligible
light in the observing ($I$) bandpass.  Note that the A and B components cannot
be identical -- that is, can not have the same masses and the same luminosities (at
least not in the I bandpass).  In such a system the center of light would track the
center of mass exactly and there would be no astrometric perturbation.  The general
equation connecting the observed semi-major axis of the photocentric orbit, $\alpha$,
with the elements of the true orbit is
\begin{displaymath}
  {\rm \alpha = a * (M_2/{(M_1+M_2)} - \beta)
         = a_1 * (1 - \beta * {(M_1+M_2)}/M_2 ) }
\end{displaymath}
where a is the semi-major axis of the relative orbit of the two components (in AU),
a$_1$ is the semi-major axis of the primary star (in AU), M$_{\rm 1}$ and M$_{\rm 2}$
are the masses of the primary and secondary components, respectively (in M$_\sun$),
and
\begin{displaymath}
  \beta = l_2/(l_1+l_2)
\end{displaymath}
is the relative luminosity of the secondary star compared with the luminosity of the
combined light (van de Kamp 1967).  If the secondary star makes a negligible contribution
to the combined light, then $\beta$ = 0 and a$_1 = \alpha$; if it contributes significantly
to the combined light, then a$_1 > \alpha$.  As for any binary system, the masses must
satisfy the general harmonic relation
\begin{displaymath}
  {\rm a{_1}{^3}/P^2 = M{_2}{^3}/(M_1+M_2)^2 }.
\end{displaymath}
where P is the period in sidereal years.

Lacking direct observational information about the relative brightnesses of the
individual A and B components at all wavelengths (and especially at I bandpass where the
perturbation was observed), we can derive constraints about possible and impossible
components from the observed orbit, the observed radial velocities, the CMDs, and
our current knowledge regarding the mass--luminosity relationships (MLRs) for low-mass
stars.  An additional potential constraint is the failure to detect a companion
in Keck II laser guide star adaptive optic observations (Siegler et al. 2007).
However, this null observation turns out to be of limited value
because the observing epoch 2006.460 was at an orbital phase (Figure 1) where
the photocenter was displaced from the barycenter by only 2 mas, implying a
separation of the pair of only 5 mas, well below their sensitivity limits.

In fact, great progress has been achieved
in the last few years in measuring dynamical masses for M dwarfs, both from ground based
observations employing various combinations of precision radial velocities, adaptive optics
imaging, and near infrared long baseline interferometry (S\'egransan et al. 2000; Delfosse
et al. 2000; Siegler et al. 2005; Siegler et al. 2007) and from space employing HST-FGS
(Benedict et al. 2001; Henry 2004).  As a result, there are now over two dozen M dwarfs
with accurately determined masses (S\'egransan et al. 2000; S\'egransan et al. 2003).
However, only one star, GJ1245C (= G208-44B = LHS3494B) with a dynamically measured mass
of 0.074$\,\pm\,$0.005 M$_{\sun}$ (Henry et al. 1999; Henry 2004), currently anchors the
region $\lesssim\,$0.10 M$_{\sun}$.  Two additional binaries -- GL569Bab, M8.5V+M9V,
(Lane et al. 2001) and 2MASSJ0746425+2000321AB, L0V+L1.5V, (Bouy et al. 2004) -- have
astrometrically determined orbits and resulting total mass determinations which indicate
individual components with masses $\lesssim\,$0.10 M$_{\sun}$.  Conversion to actual
individual masses for both systems is, unfortunately, still model dependent at this stage.
Hopefully, precision radial velocity measures will soon resolve the remaining ambiguities.
We also acknowledge in passing that many new binaries with low-mass components and estimated
periods as short as a few years have been identified recently (e.g., Forveille et al. 2005;
Siegler et al. 2005; Siegler et al. 2007) and reliable dynamical mass determinations for a
number of objects with masses $\lesssim\,$0.10 M$_{\sun}$ should be forthcoming in just
a few years.

We consider first the possibility that the secondary contributes virtually no light to
the observed combined light in the I bandpass, here taken to mean
I$_{\rm B}$$-$I$_{\rm A}$$\,\gtrsim\,$6.5 mag so that
I$_{\rm A}$$-$I$_{\rm A+B}$$\,\lesssim\,$0.01 mag.  Since all three CMDs presented in
Section 2 above suggests that the A component lies very close to the dwarf main sequence,
we estimate the mass for it from currently available empirical MLRs.  As discussed by
Delfosse et al. (2000), the current MLRs for near-infrared (JHK) bandpasses all
show small observational dispersion and agree well with the predictions of the BCAH98
models whereas the M$_{\rm V}$ versus mass MLR shows considerably more scatter and increasing
deviations from the models for masses $\lesssim\,$0.4 M$_{\sun}$.  For our initial estimate
of the mass for  LSR1610$-$00A we employ the Delfosse et al. (2000) polynomials fits to
existing observations for log[M/M$_{\sun}$] versus M$_{\rm V}$, M$_{\rm J}$, M$_{\rm H}$, and
M$_{\rm K}$.  These relations yield a mean value of 0.095~M$_{\sun}$ with a range of
0.098--0.090~M$_{\sun}$.

The first three lines of Table 4 gives the solutions for the mass of the B component,
assuming a possible range of masses for the A component as discussed above, and assuming
that $\beta$(I) = 0.  These solutions all require the companion object to have a mass
near 0.06 M$_{\sun}$.  Therefore it would certainly be a brown dwarf.  It likely would
have a spectral type of late-T or later, a temperature below 800 K, and a luminosity
below 10$^{-5.5}$ L$_{\sun}$, although these quantities depend on the uncertain age and
metallicity of the companion.  These three solutions are self-consistent, in that the low
mass required for the B component (near 0.06 M$_{\sun}$) implies that it will not contribute
significant light both in the I bandpass astrometric images and in the K bandpass
spectrophotometric data.

Also included for each solution in Table 4 are the predicted radial velocity
semi-amplitudes for the both the primary, K$_1$, and for the secondary, K$_2$,
stars.  Specifically, for the primary star we have, in km s$^{-1}$,
\begin{displaymath}
  { \rm K_1 = a_1 * sin(i) / [0.0336*P*\sqrt{(1-e^2)} ] }
\end{displaymath}
when a$_1$ is in AU and P is in sidereal years.  The predicted radial velocity
curve for the $\beta$ = 0 solutions combined with our astrometric orbital
elements is shown in Figure 6, along with the radial velocity observations 
from RB06 and this paper (see Section 2.3 above).  Given the formal error
estimates for the observations, the results seem to be self-consistent.
While better definition of the observed radial velocity curve is certainly
doable, it will require significant effort on a Keck class telescope.

Solutions were then carried out with $\beta$(I) = 0.05 and 0.10, corresponding
to a pair where
I$_{\rm B}$$-$I$_{\rm A}$$\,\approx\,$3.2 mag and 2.4 mag, respectively.
These solutions require the B component to have a higher mass in the range
0.07-0.08 M$_{\sun}$ in order to compensate for the increased semi-major axis
of the relative orbit necessary to maintain the observed semi-major axis of the
photocentric orbit.  Then it would be near
the star/brown dwarf boundary and could be either.  The absolute magnitudes
given in Table 4 for $\beta = 0.05$ and 0.10 are similar to those for
2M0532+82 (Burgasser et al. 2008), and our required mass for LSR1610$-$00B
is consistent with the mass (0.074-0.082 M$_{\sun}$) that Burgasser et al.
estimated for 2M0532+82.  One possible concern is that objects like 2M0532+82
have $I-J$ colors redder than LSR1610$-$00A, leading to the companion
contributing more light at $J$ (and probably at $H$ and $K$, depending on
its $J-K$ color).  As discussed above, evidence for a composite infrared
spectrum is not compelling.  With present available data, however,
these solutions cannot be ruled out.

Finally, solutions were attempted for a selection of $\beta$(I) greater
than 0.10.  The $\beta$(I) = 0.15 case corresponds to a pair where
I$_{\rm B}$$-$I$_{\rm A}$$\,\approx\,$1.9 mag.
Solutions with $\beta > 0.10$ appear to be incompatible with the observed
parallax and absolute magnitudes;  as $\beta$ is increased, the semi-major
axis of the primary star becomes larger, the masses of both components must
become larger to keep the orbit bound, and the high masses become inconsistent
with any reasonable mass-luminosity relation and the observed absolute
magnitudes.  Therefore, we believe these possibilities are excluded.

For all the acceptable solutions given in Table 4, the sum of the semi-major
axes of the two stars is 0.7--0.8 AU.  Allowing for the observed eccentricity
of the orbits, the separation between the two stars is 0.4--0.8 AU.
This separation is sufficiently large that mass transfer cannot be occuring now
and would not have occured in the past unless through a phase of common envelope
evolution.

\section{Galactic Orbits for LSR1425+71 and LSR1610$-$00}

The distances determined in this paper for LSR1425+71 and LSR1610$-$00, together
with radial velocities, allow calculation of their space velocities and
orbits through the galaxy.  As mentioned in Sec. 2, the large negative V
velocity for both stars is strongly indicative of a halo origin for both.
To calculate their orbits, we modeled the orbits in three potentials,
two extreme cases (a Keplerian point mass, and a Pseudo-isothermal
sphere), and a quasi-realistic model (that of Dauphole and Colin 1995,
which is made up of three Miyamoto-Nagai components, representing the
bulge, disk and halo).  The orbits were integrated using a
Bulirsh-Stoer integrator.  Following Kerr \& Lynden-Bell (1986), the
solar radius was presumed to be 8.5 kpc, and the circular velocity of
the LSR $\Theta_0 = 220.0$ km s$^{-1}$.  The solar motion with respect
to the LSR $(U,V,W) = (-10.0, 5.2, 7.2)$ km s$^{-1}$, where $U$
points radially outward (Dehnen \& Binney 1998).  Energy was conserved
at better than one part in $10^9$ over the integrations.
Integrations were run backwards in time for up to $3\times10^9$ years,
in steps of $10^5$ years.  Figures 7 and 8 show the orbits in the
Dauphole \& Colin (1995) potential for the past $5 \times 10^8$ years.

For both stars, in all cases, the orbits are quite eccentric ($e >
0.6$, and typically closer to $e\sim 0.9$), highly inclined with
respect to the galactic plane ($60\lesssim i \lesssim 84\deg$), have
periods of roughly 100 Myr, and apo-galactica near the solar orbital
radius.  In both cases, the orbits are largely radial and variably
vertical, with only a very modest amount of rotational motion in the
$x-y$ plane.

For LSR~1610$-$00, the peri-galacticon distance ranges from 140~pc in
the Dauphole \& Colin model up to 1,460~pc for a simple
pseudo-isothermal halo model (with a nice broad core).  For LSR~1425+71,
the peri-galacticon passage radii are 300~pc and 2030~pc,
respectively.  The total space velocities at peri-galacticon passages
for the Dauphole \& Colin model are 650 and 610 km s${-1}$ for
LSR1610$-$00 and LSR1425+71, respectively.  Given how close both stars
come to the galactic center,
it is quite likely that both have undergone significant scattering
within the last few orbital periods.  Given the very small peri-galacticon
distances (particularly for LSR1610$-$00), we felt it useful to assess the
impact of the massive black hole at the Galactic center on the stellar
orbits.  We added a point mass component, of $3.7 \times 10^6$ M$_{\sun}$
to the Dauphole \& Colin model.  Qualitatively, the results were the same.
The highly inclined and eccentric nature of the orbits as seen now would
argue that they are not originally disk stars, as the change in angular
momentum would be quite substantial.  Also, the fact that the orbital
apo-galactica are near the solar radius would argue that they are not
recent products of the far outer halo.

\section{Further Discussion}

Previously, both CV06 and RB06 have used extensive spectroscopic data
to conclude that LSR1610$-$00 appears peculiar (even ``schizophrenic''),
but it is most likely to be mildly metal-poor with some elements
(particularly enhanced aluminum) indicating an unusual composition.
The absolute magnitudes and colors found in this paper support the idea that
it is near solar metallicity, although
the uniquely red B$-$V color may be another indicator of unusual
composition.  However, the revised distance in this paper and the space
velocity and galactic orbit shown in Sec. 5 are paradoxical:
the space velocity and orbit are not at all disk-like, and would normally
indicate (although do not require) a lower metallicity than mildly metal-poor.
The finding in this paper of a binary companion with a separation of less
than 1 AU is a new surprise.  Using Occam's razor as a guiding principle,
we might hope to relate the peculiar abundances to some interaction or
mass transfer between the two binary components.

The enhancement of aluminum found by CV06 is probably an important clue.
Halo red giants and more massive AGB stars may produce aluminum enhancements
due to mixing burning products of the hydrogen shell to the surface.
The aluminum is produced at the expense of oxygen (Kraft 1994).  Hence, one possible
scenario that might be considered is that LSR1610$-$00B, the astrometric companion
to LSR1610$-$00A, is a white dwarf.  During the RGB or AGB phase of the evolving
white dwarf progenitor, an excess of aluminum was produced and mixed to the surface.
With a 1.7 year period, depending on the stellar mass, the red giant might be big enough
to transfer Al-enriched material to the small dwarf companion.  This
would result in the enhancement observed today.

It appears, however, that this scenario has fatal flaws and must be
discarded.  The fundamental difficulty is the low mass of both components
required by the binary possibilities in Table 4.
White dwarf masses below 0.2 M$_{\sun}$ have been found as companions to
millisecond pulsars (van Kerkwijk et al. 1996) and in the Sloan Digital
Sky Survey (Eisenstein et al. 2006), although these are rare.
What appears to be required to form a white dwarf of extremely low mass
is that the companion has a separation such that Roche lobe overflow
occurs when the primary star is trying to leave the main sequence with
only a small helium core.  If this happens, the star does not become
much bigger than its main sequence size before its evolution and growth
of the core mass are truncated.  As such, the 1.7 year period and the
separation of the two stars of 0.4-0.8 AU are too long and too large
for this to have occurred.

A second problem is that, even with a maximum cooling age on the order of
that of the halo ($\sim\,$12 Gyr), most low-mass white dwarfs will still have a
luminosity such that they will be visible in the spectrophotometry of LSR1610$-$00.
The most stringent constraint is the observed M$_{\rm B}$~=~22.4 and B$-V$~=~3.3,
requiring M$_{\rm B}\, >\,$25 for a possible white dwarf companion.
For example, white dwarf models from Bergeron et al. (1995) with hydrogen
atmospheres and M = 0.15 M$_{\sun}$ reach M$_{\rm B}$~=~19.1 at a cooling age
of 12 Gyr.  It may be possible that a low-mass white dwarf with a helium atmosphere
would cool quickly enough that it could still be present and not detected.
Otherwise, the companion to LSR1610$-$00A must be an unevolved, lower-mass star
or substellar brown dwarf.  Therefore, the B component to LSR1610$-$00A appears
not to be the source of abundance anomalies in LSR1610$-$00A.

Nevertheless, peculiar abundances do point toward the accretion of
mass onto LSR1610$-$00.  Most notable is the enhancement of Al found by CV06,
a result suggesting accretion from a massive AGB star that has undergone
hot-bottom-burning.  Some stars in globular clusters (both giants and
subgiants) have enhanced Al and Na, and depleted O, indicative of external
pollution by AGB stars (e.g., Sneden et al. 2004).
The mechanism for Al and Na production is believed to be proton capture
by Mg and Ne nuclei at high temperatures at the base of the hydrogen-burning
shell.  The most extreme abundance anomalies of this type are seen
in the extreme metal-poor star HE1327-2326 (Frebel et al. 2005) where
C, N, Na, Mg, and Al are seen to be strongly enhanced compared to Fe,
and Ca and Ti are mildly enhanced.  We speculate that accretion of
$<$0.05 M$_{\sun}$ of material with such strong enhancements onto LSR1610$-$00,
which initially might have been about 0.05 M$_{\sun}$ and [Fe/H] $\sim -$2,
followed by complete mixing in the fully convective star,
can lead to the star we see now.  The accretion enhances all elements,
making LSR1610$-$00 mildly metal-poor, it enhances C and the C/O ratio,
almost (but not quite) turning LSR1610$-$00 into a carbon star and leaving
most oxygen tied up in CO, it enhances Ti, but leaves TiO slightly
depleted, and it greatly enhances Al.  The $B$ bandpass includes several
bands of AlH, as well as MgH, SiH, and CaI:  these bands are likely
to have enhanced strength in LSR1610$-$00 compared to most dM6 and
sdM stars because of a combination of being mildly metal-poor
(strengthening the hydrides and atomic lines) and abundance anomalies.
We speculate that these (not yet observed) features are the cause
of the unique red B$-$V color.

The AGB star proposed to be the source of mass accretion probably would
have been a more distant companion in a triple system.  It would now be
a cool white dwarf, and must have separated from the present LSR1610$-$00
binary when it lost most of its mass.  (If it were still bound with LSR1610$-$00,
it would now be detectable as a companion, unless it was initially massive
enough to become a neutron star.)  Alternatively, LSR1610$-$00 and the AGB star
originally could have been part of a globular cluster when the accretion
occurred, with LSR1610$-$00 later being lost from the cluster or the cluster
dissipating entirely.

\acknowledgments

Acknowledgments...
The authors would like to express their sincere appreciation to
Elizabeth ``Betsy" Green for her advice and assistance in carrying
out the reductions of the MMT radial velocity observations.

\clearpage

\begin{deluxetable}{lr}
\tablecolumns{2}
\tablewidth{0pt}
\tablecaption{Astrometric and Photometric Results for LSR1425+71}
\tablehead{
\colhead{Result} &
\colhead{Value} }
\startdata
Epoch Range (yr)                        &             4.95  \\
No. Frames                              &               85  \\
No. Nights                              &               81  \\
No. Ref. Stars                          &               16  \\
Rel. Parallax (mas)                     & 12.27 $\pm$ 0.45  \\
Rel. Proper Motion (mas yr$^{-1}$)      &  625.8 $\pm$ 0.2  \\
P.A. of Proper Motion (degrees)         &  255.2 $\pm$ 0.1  \\
Correction to Abs. Parallax (mas)       &  1.10 $\pm$ 0.25  \\
Abs. Parallax (mas)                     & 13.37 $\pm$ 0.51  \\
V (n=4)                                 & 19.62 $\pm$ 0.02  \\
B$-$V (n=2)                             &  2.06 $\pm$ 0.05  \\
V$-$I (n=4)                             &  3.26 $\pm$ 0.02  \\
M$_{\rm{V}}$                            & 15.25 $\pm$ 0.09  \\
J$^1$                                   & 14.83 $\pm$ 0.04  \\
H$^1$                                   & 14.43 $\pm$ 0.06  \\
K$_{\rm {s}}^1$                         & 14.34 $\pm$ 0.10  \\
V$_{\rm{rad}}$ (km s$^{-1}$)$^2$      &   -65 $\pm$ 20    \\
U (km s$^{-1}$)                       &    84 $\pm$ 6     \\
V (km s$^{-1}$)                       &  -202 $\pm$ 13    \\
W (km s$^{-1}$)                       &    66 $\pm$ 14    \\
\enddata
\tablenotetext{1}{From 2MASS All-Sky Point Source Catalog}
\tablenotetext{2}{From L\'epine et al. 2003b, Table 1}
\end{deluxetable}
\clearpage

\begin{deluxetable}{lr}
\tablecolumns{2}
\tablewidth{0pt}
\tablecaption{Astrometric and Photometric Results for LSR1610$-$00}
\tablehead{
\colhead{Result} &
\colhead{Value} }
\startdata
Epoch Range (yr)                        &             4.27  \\
No. Frames                              &              219  \\
No. Nights                              &              139  \\
No. Ref. Stars                          &               15  \\
Rel. Parallax (mas)$^1$                 & 30.74 $\pm$ 0.52  \\
Rel. Parallax (mas)$^2$                 & 30.02 $\pm$ 0.21  \\
Rel. Proper Motion (mas yr$^{-1}$)$^1$  & 1447.2 $\pm$ 0.3  \\
Rel. Proper Motion (mas yr$^{-1}$)$^2$  & 1446.5 $\pm$ 0.2  \\
P.A. of Proper Motion (degrees)$^1$     &  213.4 $\pm$ 0.1  \\
P.A. of Proper Motion (degrees)$^2$     &  213.3 $\pm$ 0.1  \\
Correction to Abs. Parallax (mas)       &  1.00 $\pm$ 0.15  \\
Abs. Parallax (mas)$^2$                 & 31.02 $\pm$ 0.26  \\
V (n=4)                                 & 19.10 $\pm$ 0.02  \\
B$-$V (n=2)                             &  3.26 $\pm$ 0.14  \\
V$-$I (n=4)                             &  4.05 $\pm$ 0.02  \\
M$_{\rm{V}}$(A+B)                       & 16.56 $\pm$ 0.03  \\
J$^3$                                   & 12.91 $\pm$ 0.02  \\
H$^3$                                   & 12.32 $\pm$ 0.02  \\
K$_{\rm {s}}^3$                         & 12.02 $\pm$ 0.03  \\
V$_{\rm{rad}}$ (km s$^{-1}$)$^4$      & -101.0 $\pm$ 1.0  \\
U (km s$^{-1}$)                       &     36 $\pm$ 2    \\
V (km s$^{-1}$)                       &   -232 $\pm$ 2    \\
W (km s$^{-1}$)                       &    -61 $\pm$ 2    \\
\enddata
\tablenotetext{1}{Before removal of the astrometric perturbation}
\tablenotetext{2}{After removal of the astrometric perturbation}
\tablenotetext{3}{From 2MASS All-Sky Point Source Catalog}
\tablenotetext{4}{Adopted systemic radial velocity (see Sec. 4)}
\end{deluxetable}
\clearpage

\begin{deluxetable}{lr}
\tablecolumns{2}
\tablewidth{0pt}
\tablecaption{Photocentric Orbital Elements for LSR1610$-$00AB}
\tablehead{
\colhead{Parameter} &
\colhead{Value} }
\startdata
Period (yr)    &    1.662  $\pm$   0.012 \\
$\alpha$ (mas)$^1$&  8.91  $\pm$   0.31  \\
$\alpha$ (AU)$^1$ & 0.276  $\pm$   0.010 \\
i (deg)        &     83.2  $\pm$   1.0   \\
e              &    0.444  $\pm$   0.017 \\
$\omega$ (deg) &    151.4  $\pm$   4.6   \\
$\Omega$ (deg) &    102.8  $\pm$   1.8   \\
T$_0$          &  2005.854 $\pm$   0.020 \\
RMS$_{\rm RA}$ (mas)  &  2.2 \\
RMS$_{\rm Dec}$ (mas) &  3.3 \\
\enddata
\tablenotetext{1}{Semi-major axis of the photocentric orbit.}
\end{deluxetable}
\clearpage

\begin{deluxetable}{llrcrrrrcrr}
\tablecolumns{11}
\tablewidth{0pt}
\tablecaption{Viable LSR1610$-$00AB Component Properties}
\tablehead{
\colhead{} &
\multicolumn{5}{l}{Primary Star} &
\multicolumn{5}{l}{Secondary Star} \\
\colhead{$\beta$} &
\colhead{M$_1$} &
\colhead{a$_1$} &
\colhead{K$_1$} &
\colhead{$M{_I}$} &
\colhead{$M{_K}$} &
\colhead{M$_2$} &
\colhead{a$_2$} &
\colhead{K$_2$} &
\colhead{$M{_I}$} &
\colhead{$M{_K}$} \\
\colhead{} &
\colhead{(M$_{\sun}$)} &
\colhead{(AU)} &
\colhead{(km s$^{-1}$)} &
\colhead{} &
\colhead{} &
\colhead{(M$_{\sun}$)} &
\colhead{(AU)} &
\colhead{(km s$^{-1}$)} &
\colhead{} &
\colhead{} }
\startdata
  0.00& 0.090& 0.287& 5.7 & 12.5 & 9.5 & 0.057 & 0.453 & 9.0 &$>$18.&$>$15. \\
  0.00& 0.095& 0.287& 5.7 & 12.5 & 9.5 & 0.059 & 0.462 & 9.2 &$>$18.&$>$15. \\
  0.00& 0.100& 0.287& 5.7 & 12.5 & 9.5 & 0.061 & 0.472 & 9.4 &$>$18.&$>$15. \\
  0.05& 0.095& 0.325& 6.5 & 12.5 & 9.5 & 0.070 & 0.441 & 8.8 & 15.7 & 12.7 \\
  0.10& 0.095& 0.366& 7.3 & 12.5 & 9.5 & 0.082 & 0.424 & 8.5 & 14.9 & 11.9 \\
\enddata
\end{deluxetable}
\clearpage

\begin{figure}
\plotone{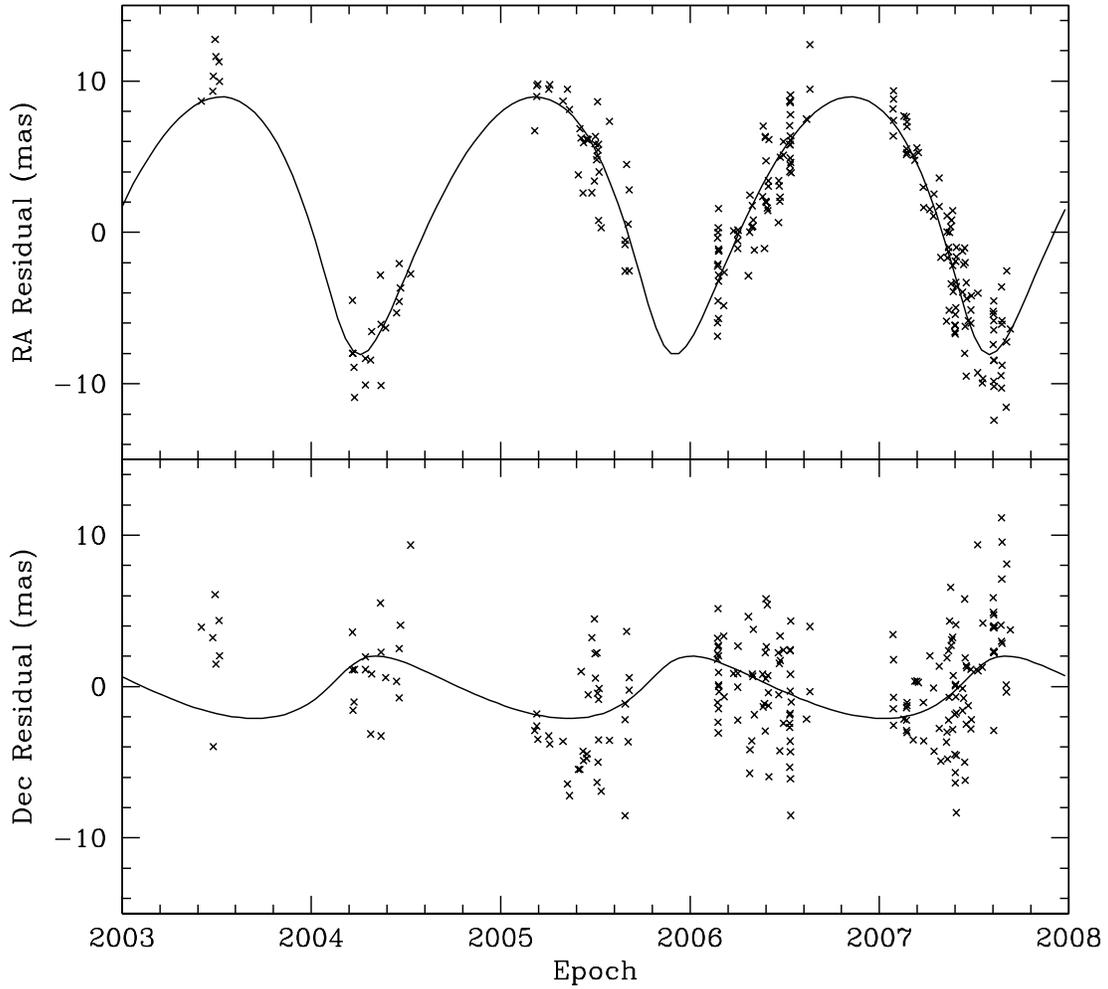}
\caption{The positions in Right Ascension and Declination for the LSR1610$-$00AB
photocenter after removing the parallactic and proper motions.  The curves show
the orbital motion fit to these residuals and represented by the orbital elements
given in Table 3.
\label{Fig.1}}
\end{figure}
\clearpage
 
\begin{figure}
\plotone{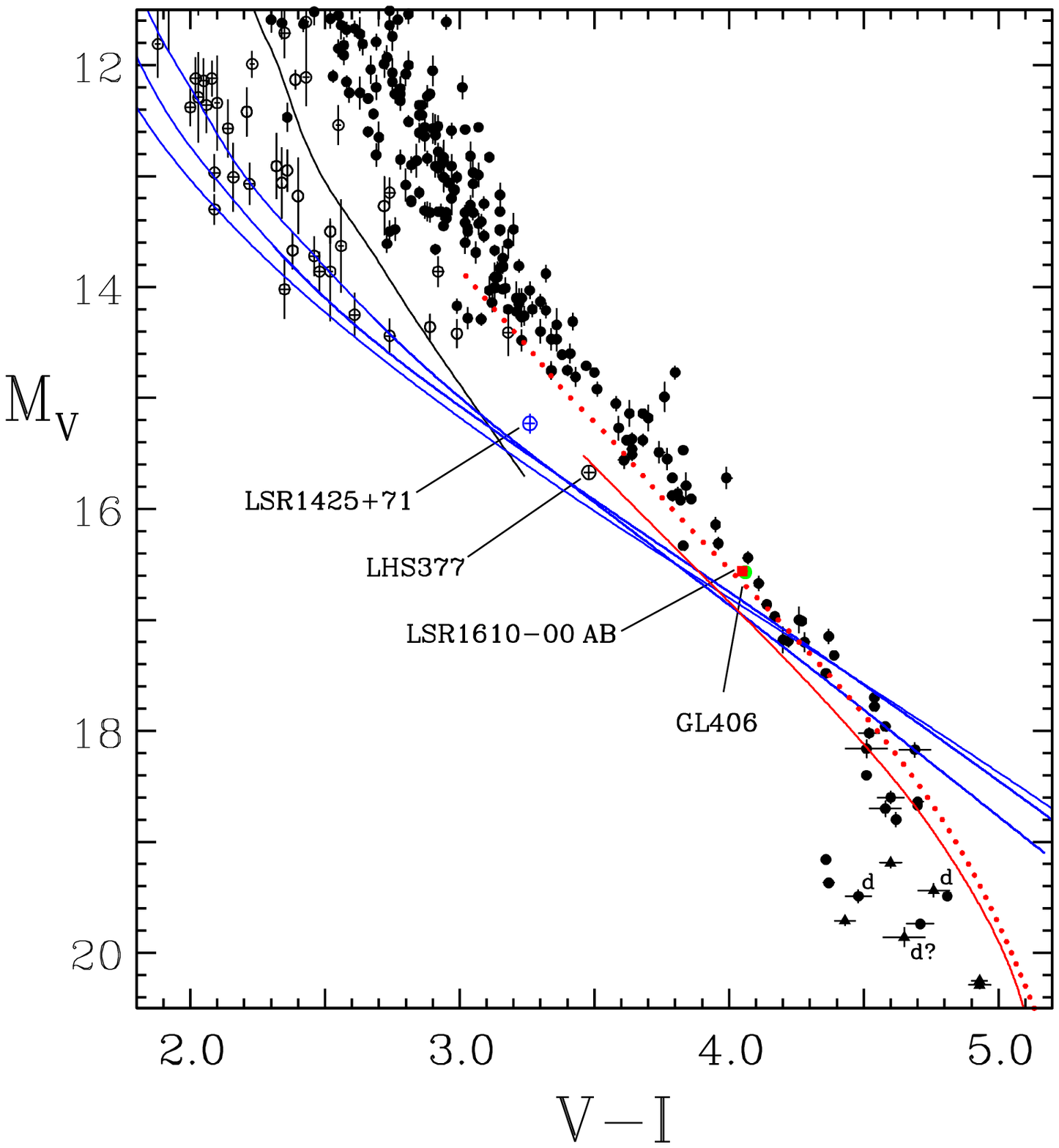}
\caption{LSR1425+71 and LSR1610$-$00 in the M$_{\rm V}$ versus V$-$I diagram.
A selection of M-type dwarfs (solid circles), M-type subdwarfs (open circles),
and early L-type dwarfs (solid triangles) are plotted for reference and comparison.
The solid blue lines are the Baraffe et al. (1997) isochrones for metallicities of
[M/H] =  $-$1.0, $-$1.5, and $-$2.0 for an age of 10 Gyrs.  The solid black 
line is the isochrone for solar metallicity stars with masses $\geq\,$0.10~M$_{\sun}$
and with ages of 0.5~Gyr from Baraffe et al. (1998).  The solid red and dotted red
lines represent the isochrones for DUSTY models from Chabrier et al. (2000) with
masses $\leq\,$0.10~M$_{\sun}$ and with ages of 0.5~Gyr and 0.1~Gyr, respectively.
Three of the latest dwarfs plotted that are either known or suspected double stars
are flagged with `d' or `d?'.  The two other labeled stars (LHS377 and GL406) are
discussed in the text.
\label{Fig.2}}
\end{figure}
\clearpage

\begin{figure}
\plotone{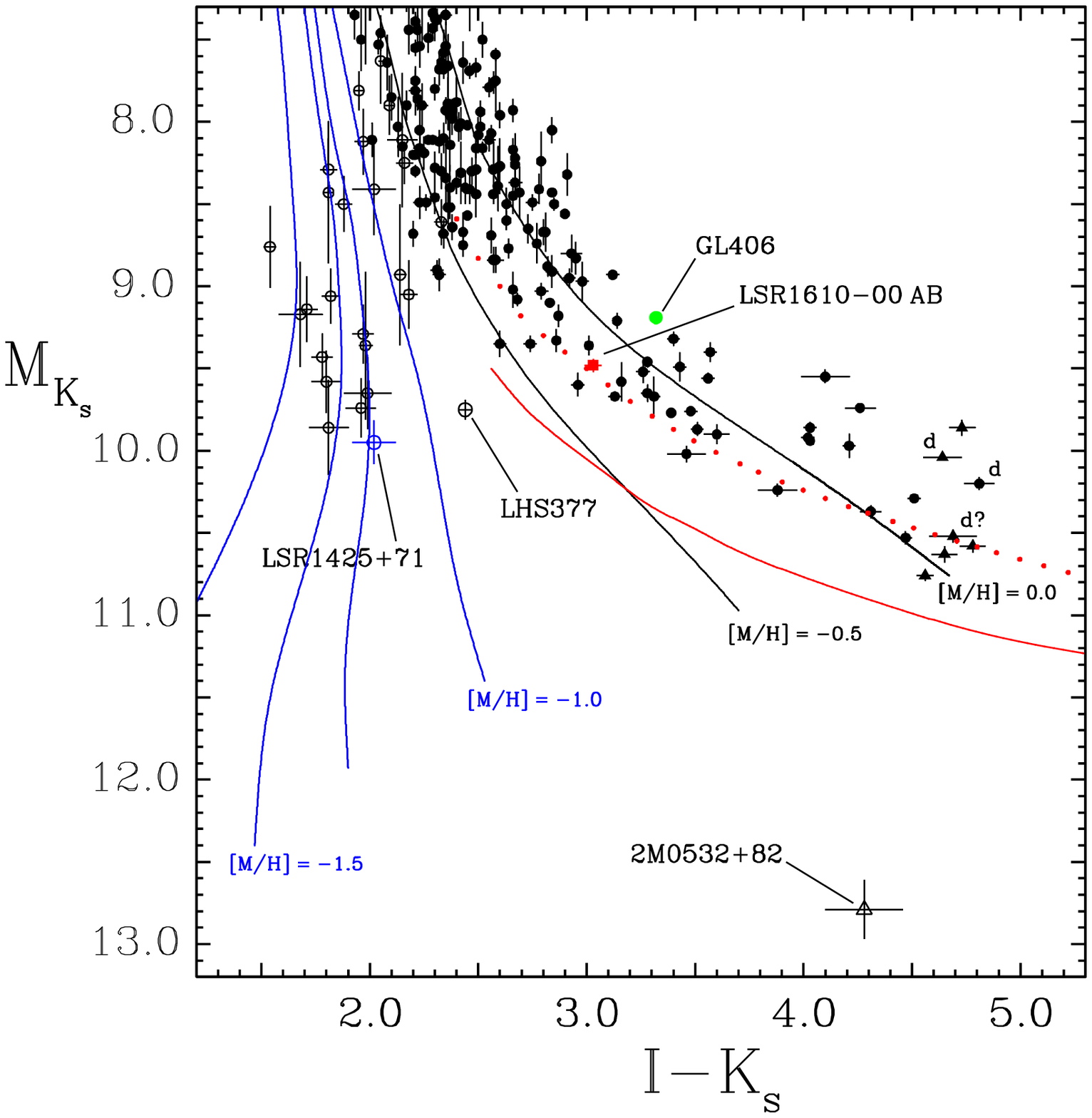}
\caption{LRS1425+71 and LSR1610$-$00 in the M$_{\rm K_{\rm s}}$ versus
I$-$K$_{\rm s}$ diagram.
A selection of M-type dwarfs (solid circles), M-type subdwarfs (open circles),
and early L-type dwarfs (solid triangles) are plotted for reference and comparison.
The solid blue lines are the Baraffe et al. (1997) isochrones for metallicities of
[M/H] =  $-$1.0, $-$1.3, $-$1.5, and $-$2.0 for an age of 10 Gyrs.
The solid black lines are isochrones for solar metallicity and slightly metal-poor
stars with masses $\geq\,$0.10~M$_{\sun}$ and with ages of 0.5~Gyr from Baraffe
et al. (1998).
The solid red and dotted red
lines represent the isochrones for DUSTY models from Chabrier et al. (2000) with
masses $\leq\,$0.10~M$_{\sun}$ and with ages of 0.5~Gyr and 0.1~Gyr, respectively.
Three of the latest dwarfs plotted that are either known or suspected double stars
are flagged with `d' or `d?'.  The three other labeled stars (LHS377, GL406, and
2M0532+82) are discussed in the text.
\label{Fig.3}}
\end{figure}
\clearpage

\begin{figure}
\plotone{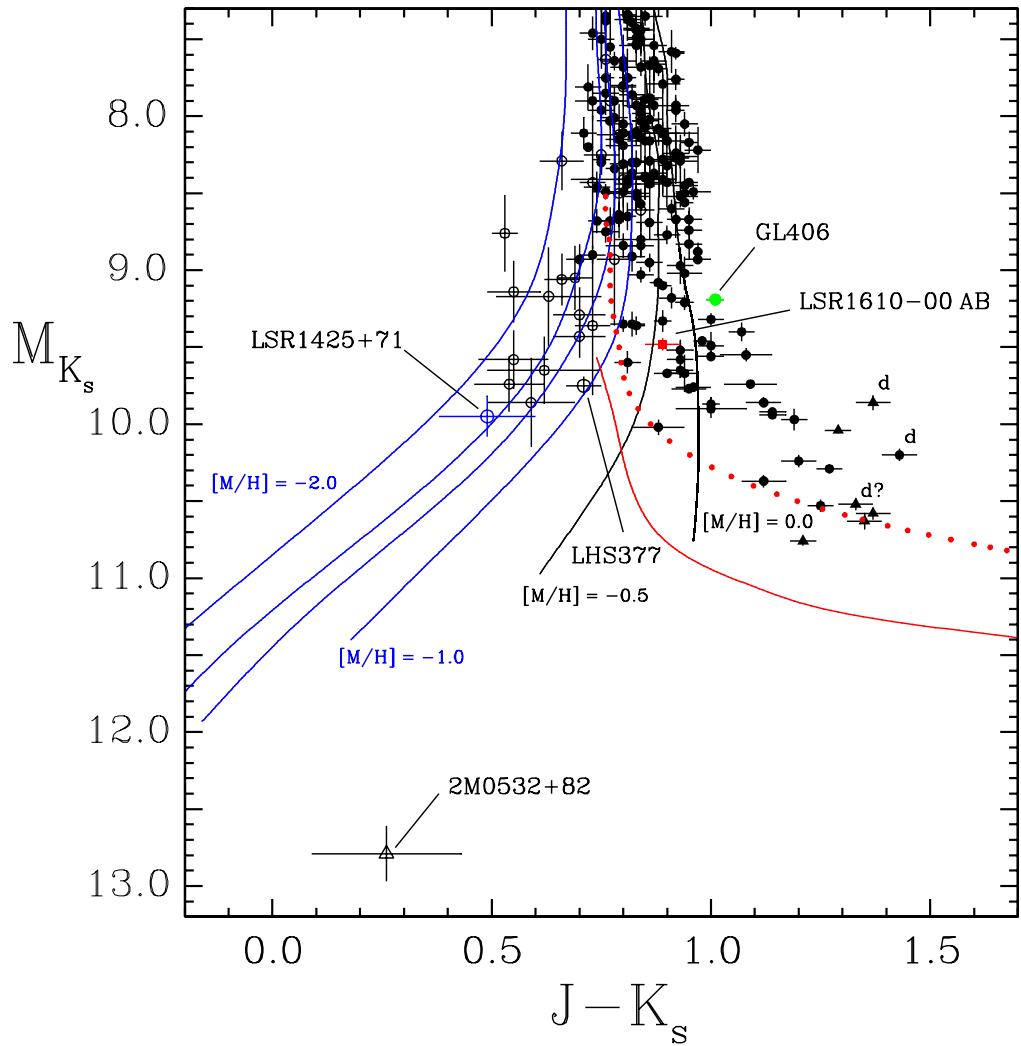}
\caption{LRS1425+71 and LSR1610$-$00 in the M$_{\rm K_{\rm s}}$ versus
J$-$K$_{\rm s}$ diagram.  The symbols and isochrones shown are the same as in
Fig. 3.
\label{Fig.4}}
\end{figure}
\clearpage

\begin{figure}
\plotone{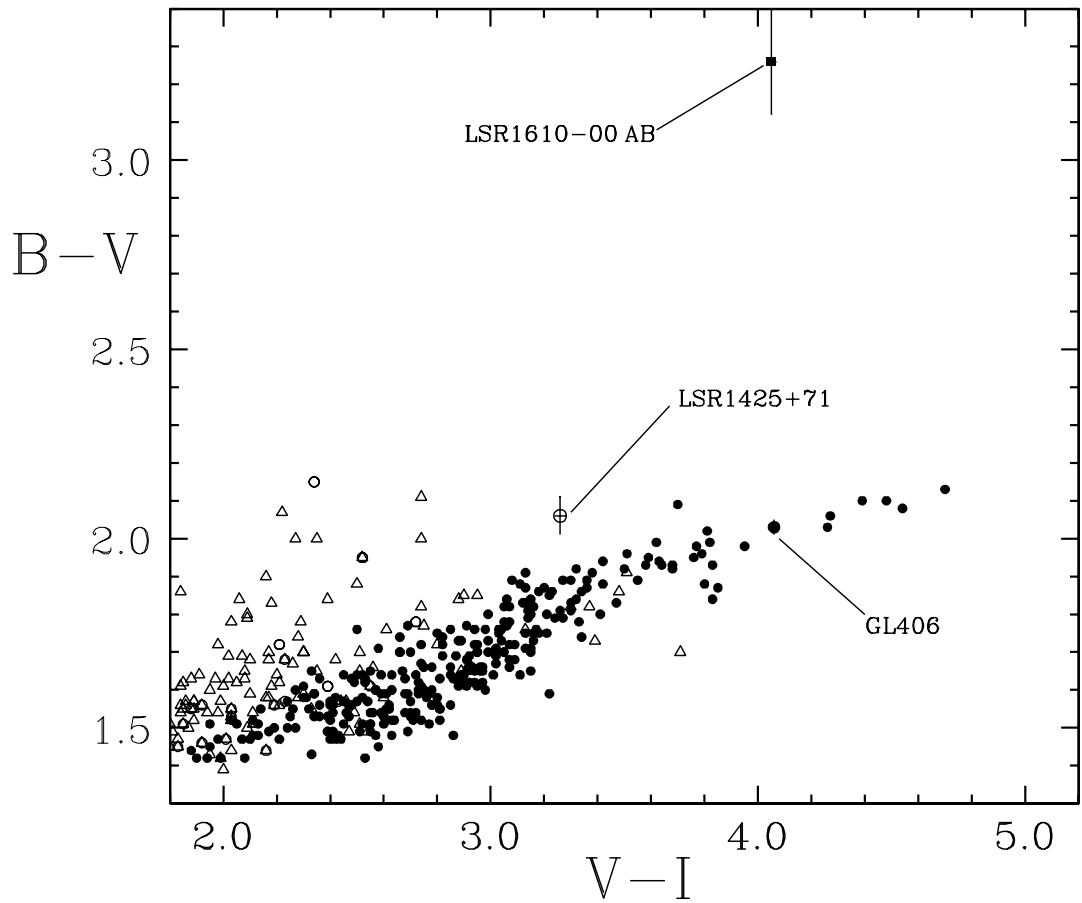}
\caption{LSR1425+71, LSR1610$-$00AB, and GL406 in the B$-$V versus V$-$I diagram.
Solid circles and open circles are M-type dwarfs and subdwarfs, respectively, as in
Figure 2.  Open triangles are stars spectroscopically classified as subdwarfs but
lacking reliable parallax distances at this time.
\label{Fig.5}}
\end{figure}
\clearpage

\begin{figure}
\plotone{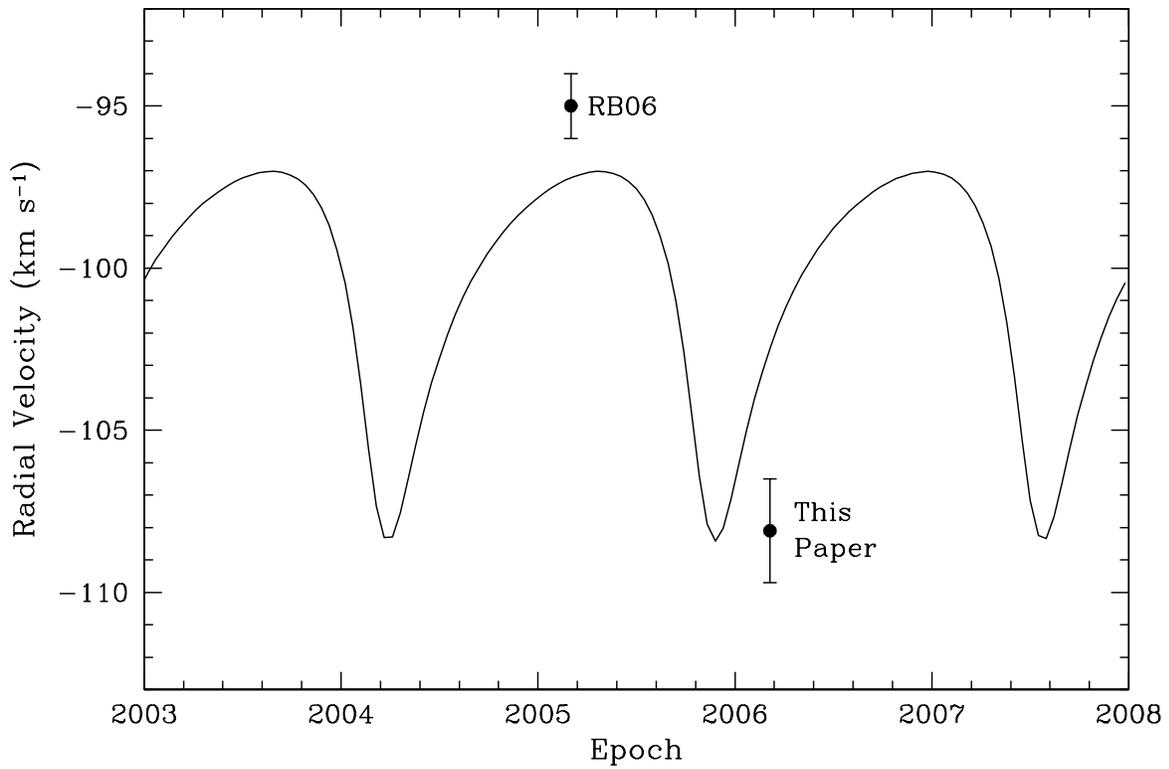}
\caption{Heliocentric radial velocity observations of LSR1610$-$00 from RB06 and
this paper, together with the orbital velocity curve predicted by the astrometric
orbit with $\beta = 0$ (see text).  The one other radial velocity observation by
LRS03 has a large error and is not plotted.
\label{Fig.6}}
\end{figure}
\clearpage

\begin{figure}
\plotone{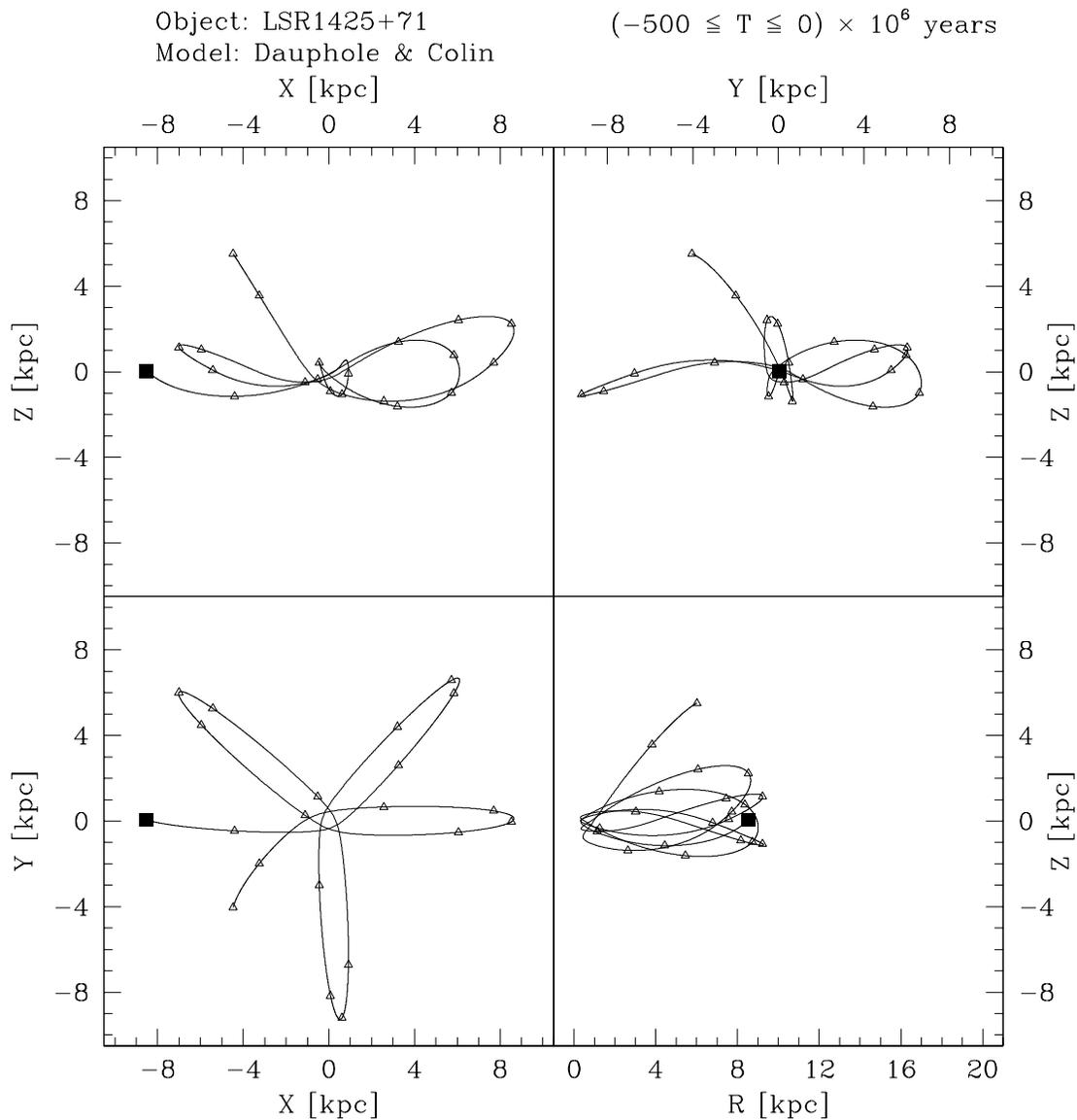}
\caption{The $x-y$, $x-z$, $y-z$ and $R-z$ projections of the
orbit of LSR1425+71 integrated backwards for
$5\times 10^8$ years in the Galactic model potential of Dauphole
\& Colin (1995).  The filled square marks the star's position now (and
given the resolution, it also indicates the location of the solar
neighborhood).  The open triangles are placed at $25 \times 10^6$ year
intervals.
\label{Fig.7}}
\end{figure}
\clearpage

\begin{figure}
\plotone{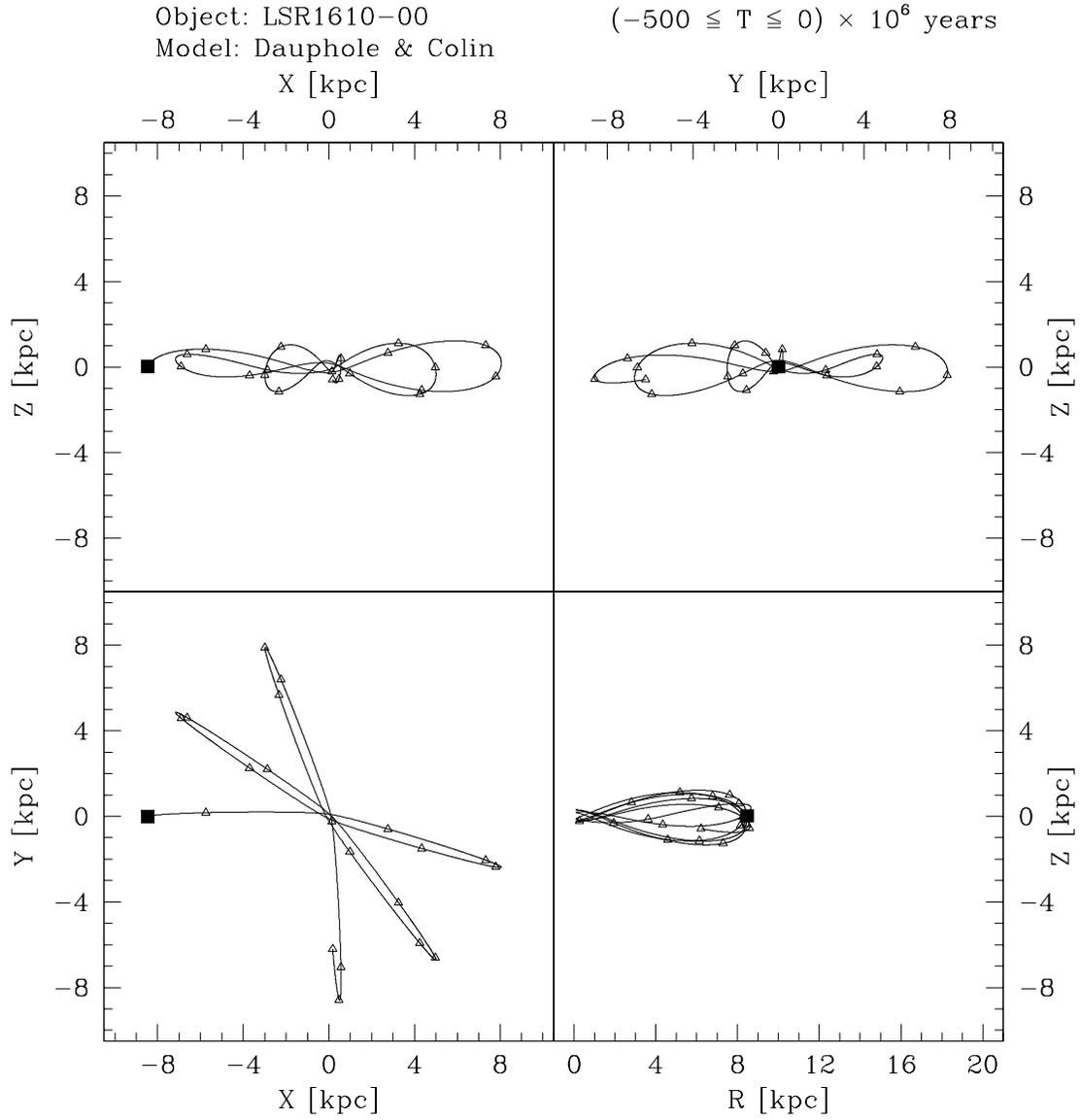}
\caption{Projections of the orbit of LSR1610$-$00, where the panels and the
symbols are the same as in Fig. 7.
\label{Fig.8}}
\end{figure}

\end{document}